\documentclass[preprint, 5p, twocolumn, number, sort&compress, pdftex]{elsarticle}
\usepackage[T1]{fontenc}
\usepackage[latin1]{inputenc}
\usepackage[english]{babel}
\usepackage{amssymb, amsmath}
\usepackage{bm}
\usepackage{textcomp}
\usepackage{color}
\usepackage{lmodern}
\usepackage[bookmarksnumbered=true, bookmarksopen=false, colorlinks]{hyperref}

\hypersetup{
	pdfauthor  = "Pocsai Mihály András",
	pdftitle   = "Electron Acceleration by a Bichromatic Chirped Laser Pulse in Underdense Plasmas",
	pdfsubject = "Applied Physics"
}

\newcommand{\parenth}[1]{\left( #1 \right)}
\newcommand{\bparenth}[1]{\left[ #1 \right]}

\newcommand{\abs}[1]{\left\lvert #1 \right\rvert}

\newcommand{\diff}{\mathrm{d}}

\journal{Nuclear Instruments and Methods in Physics Research B}

\begin{document}
\title{Electron acceleration by a bichromatic chirped laser pulse in underdense plasmas}

\begin{frontmatter}
\author[wigner]{M.A.~Pocsai\corref{pocsai}}
\cortext[pocsai]{Corresponding author}
\ead{pocsai.mihaly@wigner.mta.hu}

\author[wigner,eli]{S.~Varró}
\ead{varro.sandor@wigner.mta.hu}

\author[wigner,eli]{I.F.~Barna}
\ead{barna.imre@wigner.mta.hu}

\address[wigner]{Wigner Research Centre for Physics of the Hungarian Academy of Sciences, Konkoly--Thege Miklós út 29-33, H-1121 Budapest, Hungary}
\address[eli]{ELI-HU Nonprofit Ltd.,	Dugonics Tér 13, H-6720 Szeged, Hungary}

\begin{abstract}
A theoretical study of laser and plasma based electron acceleration is presented. An effective model has been used, in which the presence of an underdense plasma has been taken account via its index of refraction $n_{m}$. In the confines of this model, the basic phenomena can be studied by numerically solving the classical relativistic equations of motion. The key idea of this paper is the application of chirped, bichromatic laser fields. We investigated the advantages and disadvantages of mixing the second harmonic to the original $\lambda = 800 \, \mathrm{nm}$ wavelength pulse. We performed calculations both for plane wave and Gaussian pulses.
\end{abstract}

\end{frontmatter}

\section{Introduction}
The history of laser--plasma based electron acceleration began in the late '70-s. Tajima and Dawson predicted that the plasma wakes, generated by the ponderomotive force of short, intense laser pulses, are capable to accelerate bunches of electrons effectively \citep{Tajima}. In this scheme, there are many methods for generating high amplitude plasma wakes, for a summary, see Ref.~\citep{Esarey2}. The key point in every method is the resonant excitation of the plasma. In the middle '80-s, the invention of the Chirped Pulse Amplification (CPA) made it possible to generate short, intense laser pulses without damaging the medium \cite{cpa2}. This new technology is an important milestone both in the history of laser physics and compact, plasma based particle accelerators. Nowadays, thanks to the advanced technological developments, the vision of Tajima and Dawson is getting reality.

There is a serious demand for that, since the conventional storage ring technology has reached its limits: the amplitude of the accelerating gradient must not be larger than $50 \, \mathrm{MV}/\mathrm{m}$. The violation of this criterion would lead to electric discharges that would critically damage the system. The present state-of-the-art technology, namely, the CERN--LHC, is based on $8.3 \, \mathrm{T}$ strong superconducting magnets. This way, the theoretical maximum of center-of-mass (CM) energy is $14 \, \mathrm{TeV}$. According to the most recent news, the LHC is operating at $13 \, \mathrm{TeV}$ at the moment. Using the standard technology, the maximal CM energy can be improved in two independent ways. One either has to build a larger ring (VLHC), with a circumference of $80$ or $100 \, \mathrm{km}$s, or develop stronger (e.g.~$21 \, \mathrm{T}$) superconducting magnets. Both solutions would be extremely expensive, and the latter is also very uncertain, there is no guarantee for it to succeed. Due to theses difficulties, new technologies are needed. The most popular of them is the concept of laser--plasma based particle accelerators, that has been mentioned above. Nowadays, there are promising experimental results for building compact particle accelerators: electrons have been accelerated up to multiples of $\mathrm{GeV}$s within a few $\mathrm{cm}$ long plasma cell \cite{leemans_multi-gev_2014, malka_electron_2002}. CERN is also open for new technologies: the construction of the CERN AWAKE experiment (Proton Driven Plasma Wakefield Acceleration) has already begun. The details of the proposed scheme can be found in \cite{pdpwfa,Caldwell}. It is important to mention that this scheme has been designed for electron acceleration only. For a long time, it was thought that acceleration of positrons by plasma wakes is impossible. Recently, it has been shown that by applying a ``doughnut shaped'' driver pulse, laser--plasma based positron acceleration can be realized as well \cite{Vieira}.

This study is the sequel of our recent work \cite{Pocsai-LPB}. In that paper, an effective theory for describing electron acceleration in underdense plasmas has been presented. We showed that a single electron can be effectively accelerated both by monochromatic planewave pulses and Gaussian laser pulses, up to $270 \, \mathrm{MeV}$, which agrees quite well with other theoretical and experimental results. In the present paper we investigate the advantages of applying a bichromatic driver pulse, namely, adding the second harmonic to the original laser pulse. Ehlotzky's work \cite{ehlotzky_bichromatic_2001}, which summarizes various relevant atomic phenomena in bichromatic laser fields gave us a good motivation and serves as a starting point of our present work. To our knowledge, there is no such approach in the literature. However, there is a nice proposal for producing narrow-energy-spread electron bunches from laser wakefield acceleration, using bichromatic laser pulses \cite{bichromatic}.

In Section \ref{sec:theory}, the fundamental theoretical basics of laser--plasma based electron acceleration are overviewed, within the confines of our effective theory. The results are presented in Section \ref{sec:results}. Finally, we summarize our work.

\section{Theory}\label{sec:theory}
During the laser--electron interaction, the Lorentz-force drives the motion of the electron:
\begin{equation}
	\mathbf{F} = e\parenth{\mathbf{E} + \mathbf{v} \times \mathbf{B}}
\end{equation}
with $e$ the electron charge, $\mathbf{E}$ the electric field, $\mathbf{B}$ the magnetic field, $\mathbf{v}$ the velocity of the electron and $\mathbf{F}$ the Lorentz-force. At sufficiently high intensities, the electron becomes relativistic. Therefore, one has to solve the relativistic Newton--Lorentz equation:
\begin{subequations}\label{eqgrp:Newton--Lorentz}
\begin{align}
	\frac{\diff \mathbf{p}}{\diff t}& = e\parenth{\mathbf{E} + \frac{\mathbf{p}}{m_{e} \gamma} \times \mathbf{B}},\\
	\frac{\diff \gamma}{\diff t}& = \frac{1}{m_{e} c^{2}} \mathbf{F} \cdot \mathbf{v}.
\end{align}
\end{subequations}
It is known that the electromagnetic field has to satisfy the electromagnetic wave equation. This condition yields the most general form for the electric and magnetic field:
\begin{align}
	\mathbf{E}(t, \mathbf{r})& = \pmb{\varepsilon} E_{0} f \bparenth{\omega \Theta \parenth{t, \mathbf{r}}},\label{eq:EM_general_E}\\
	\mathbf{B}(t, \mathbf{r})& = \frac{1}{c} \mathbf{n} \times \mathbf{E}(t)\label{eq:EM_general_B}
\end{align}
with $\pmb{\varepsilon}$ the polarization vector, $E_{0}$, the amplitude of the electric field, $\omega$ the angular frequency and $\mathbf{n}$ the unit vector of the propagation of the electromagnetic field, respectively. $f$ may be any arbitrary, smooth function. For a better transparency, we introduced the following notation, since the electromagnetic field depends only on the planewave-argument
\begin{equation}\label{eq:Theta}
	\Theta(t, \mathbf{r}) := t - \mathbf{n} \cdot \frac{\mathbf{r}}{c}.
\end{equation}

If we also want to take into account the presence of a medium with an index of refraction $n_{m} < 1$---for instance, an underdense plasma, which is the field of our present investigation---, we need to generalize the definition of $\Theta(t, \mathbf{r})$ in the following way:
\begin{equation}\label{eq:Theta-gen}
	\Theta(t, \mathbf{r}, n_{m}) := t - n_{m} \mathbf{n} \cdot \frac{\mathbf{r}}{c}.
\end{equation}
We interpret the generalized definition of $\Theta$ such that it describes the electron propagation in an underdense plasma \cite{VarroNm}. All the background effects are incorporated into $n_{m}$, which depends on the laser and plasma frequencies $\omega_{L}$ and $\omega_{p}$, respectively, in the following way:
\begin{equation}
	n_{m} = \sqrt{1 - \frac{\omega_{p}^{2}}{\omega_{L}^{2}}}
\end{equation}
with
\begin{equation}
	\omega_{p}^{2} = \frac{n_{e}e^{2}}{\varepsilon_{0}m_{e}}
\end{equation}
with $n_{e}$ the electron density in the plasma, $\varepsilon_{0}$ the permittivity of vacuum and $m_{e}$ the electron mass.

This approximation yields an effective theory which is useful for performing basic studies of the laser and plasma based electron acceleration via numerically solving the relativistic equations of motion \eqref{eqgrp:Newton--Lorentz}, treating $\mathbf{E}$ and $\mathbf{B}$ as a function of $\Theta(t, \mathbf{r}, n_{m})$. The $n_{m}=1$ case describes the pure laser based electron acceleration. It is important to mention that a single electron cannot gain a net energy from a plane wave pulse in vacuum since during one pulse period, the electron gains and loses the same amount of energy during oscillating in the laser field. However, this symmetry can be broken by applying a (linear) chirp to the laser frequency:
\begin{equation}
	\omega(t) = \omega_{0} + \sigma t,
\end{equation}
$\sigma$ being the chirp parameter and $\omega_{0}$ the initial frequency of the laser.

It is convenient and useful to rescale the parameters and introduce the following dimensionless variables:
\begin{gather}
a_{0} = \frac{e E_{0}}{m_{e} \omega_{0} c}, \nonumber\\
\omega_{0} t \rightarrow t, \quad \frac{\omega_{0}}{c} \mathbf{r} \rightarrow \mathbf{r}, \quad \frac{\mathbf{p}}{m_{e} c} \rightarrow \mathbf{p},\label{eqs:dimensionless}\\
\omega_{0} \Theta \rightarrow \Theta, \quad \frac{\sigma}{\omega_{0}^{2}} \rightarrow \sigma. \nonumber
\end{gather}
The equations of motion, expressed in terms of these new variables, take the following form:
\begin{subequations}\label{eqgrp:Newton--Lorentz_dimensionless}
\begin{align}
	\frac{\diff \mathbf{p}}{\diff t}& = a_{0} \parenth{\mathbf{E} + \frac{\mathbf{p}}{\gamma} \times \mathbf{B}},\\
	\frac{\diff \gamma}{\diff t}& = \frac{a_{0}}{\gamma} \mathbf{E} \cdot \mathbf{p}.
\end{align}
\end{subequations}

In general, a bichromatic electromagnetic field can be expressed in the compact form of
\begin{subequations}\label{eqgrp:bichromatic}
\begin{align}
	\mathbf{E} &= \mathbf{E}_{1} + \frac{A}{q} \mathbf{E}_{q},\\
	\mathbf{B} &= \mathbf{B}_{1} + \frac{A}{q} \mathbf{B}_{q}
\end{align}
\end{subequations}
with $\mathbf{E}_{1}$ and $\mathbf{B}_{1}$ being the electric and magnetic fields of the main harmonic and $0 \leq A \leq 1$ the relative amplitude of the harmonics. $q$ denotes the index of the (higher) harmonic with $q \omega_{0}$ initial frequency. The $q^{-1}$ factor is a direct consequence of the definition of the intensity parameter $a_{0}$ (see Eq.~\eqref{eqs:dimensionless}):
\begin{equation}
	a_{0} (q \omega_{0}) = \frac{a_{0} (\omega_{0})}{q}
\end{equation}

For first, we define the $q^{th}$ harmonic of a bichromatic plane wave pulse with a sine-square shaped temporal envelope:
\begin{equation}\label{eq:planewave}
	f \parenth{\Theta} = \left\lbrace \begin{array}{ll}
		\sin^{2} \parenth{\frac{\pi \Theta}{\omega_{0} T}} \times &\\
		\sin \parenth{q \Theta + q^{2} \sigma_{q} \Theta^{2} + \varphi_{q}}& \textrm{if $\Theta \in \bparenth{0, T}$}\\
		0& \textrm{otherwise}
	\end{array}
	\right.
\end{equation}
with $T$ the pulse duration, $\sigma_{q}$ the dimensionless chirp parameter and $\varphi_{q}$ with $q=1$ the carrier--envelope phase and $\varphi_{q}$ with $q > 1$ the relative phase of the harmonics. The electric field is polarized in the $x$ direction and propagates in the $y$ direction, that is, $\pmb{\varepsilon} = \mathbf{e}_{x}$, $\mathbf{n} = \mathbf{e}_{y}$.

Gaussian pulse shapes provide a more realistic description of laser beams. The mathematical expressions for an $x$-polarized Gaussian beam that propagates in the $z$ direction can be derived from the paraxial approximation \cite{Davis, Lax}. The explicit expressions containing first order corrections can be found in many works, see e.g.~\cite{Sohbatzadeh1, Sohbatzadeh2}. Here we only present the generalizations of these formulae for the $q^{th}$ harmonic in terms of dimensionless variables:

\begin{subequations}\label{eqgrp:Gauss_E}
\begin{align}
	\begin{split}
	E_{q, x}& = \frac{W_{0}}{W_{q}(z)} \exp \bparenth{-\frac{r^{2}}{W_{q}^{2}(z)}} \exp \bparenth{-\frac{\Theta^{2}}{\parenth{\omega_{0} T}^{2}}} \times \\
	& \quad \cos \bparenth{\frac{k_{q} r^{2}}{2 R_{q}(z)} - \Phi_{q}(z) + q \Theta + q^{2} \sigma \Theta^{2} + \varphi_{q}},
	\end{split}\\
	E_{q, y}& = 0,\\
	\begin{split}
	E_{q, z}& = -\frac{x}{R_{q} (z)}E_{q, x} +\\
	& \quad \frac{2x}{k_{q} W_{q}^{2}(z)} \cdot \frac{W_{0}}{W_{q}(z)} \exp \bparenth{-\frac{r^{2}}{W_{q}(z)}} \exp \bparenth{-\frac{\Theta^{2}}{\parenth{\omega_{0} T}^{2}}} \times \\
	& \quad \sin \bparenth{\frac{k_{q}r^{2}}{2 R_{q}(z)} - \Phi_{q}(z) + q \Theta + q^{2} \sigma \Theta_{q}^{2} + \varphi_{q}}
	\end{split}
\end{align}
\end{subequations}
and the magnetic field is given by
\begin{subequations}\label{eqgrp:Gauss_B}
\begin{align}
	B_{q, x}& = 0,\\
	B_{q, y}& = E_{q, x},\\
	\begin{split}
	B_{q, z}& = \frac{y}{R_{q}(z)}E_{q, x} +\\
		&\frac{2y}{k_{q} W_{q}^{2}(z)} \cdot \frac{W_{0}}{W_{q}(z)} \exp \bparenth{-\frac{r^{2}}{W_{q}^{2}(z)}} \exp \bparenth{-\frac{\Theta^{2}}{\parenth{\omega_{0} T}^{2}}} \times \\
	& \quad \sin \bparenth{\frac{k_{q}r^{2}}{2 R_{q}(z)} - \Phi_{q}(z) + q \Theta + q^{2} \sigma \Theta^{2} + \phi_{q}}
	\end{split}
\end{align}
\end{subequations}
with $W_{0} = (\omega_{0}/c) \sqrt{\lambda_{0} z_{q, R} / \pi}$ the beam waist, $W_{q}(z) = (\omega_{0}/c) \bparenth{1 + (z/z_{q, R})^{2}}^{1/2}$ the beam radius at distance $z$, $R_{q}(z) = (\omega_{0}/c) z \bparenth{1 + (z_{q, R}/z)^{2}}$ the radius of curvature, $\Phi_{q}(z) = \tan^{-1} (z/z_{q, R})$ the Guoy phase, $z_{q, R}$ the Rayleigh length, $k_{q} = q (1 + q \sigma_{q} \Theta)$ the dimensionless wavenumber of the $q^{th}$ harmonic, $\lambda_{0}$ the initial wavelength of the main harmonic and $T$ the pulse duration.

The construction of a bichromatic Gaussian pulse implies some problems. Theoretically, there is a freedom by defining the parameters of the individual harmonics: one has to choose either the beam waists or the Rayleigh lengths to be equal. Since the beam waist and the Rayleigh length depend on each other, equal Rayleigh lengths would result in different beam waists. From the expressions given above if follows that the main and the higher harmonics have the same beam waist, this is the physically relevant choice. However, the equality of the beam waists implies the difference of the Rayleigh lengths. It is also clear from equations \eqref{eqgrp:Gauss_E}, \eqref{eqgrp:Gauss_B} and \eqref{eqgrp:bichromatic} that the precise mathematical expression of a bichromatic Gaussian pulse is rather complicated.

\section{Results}\label{sec:results}
In this section we discuss the advantages and disadvantages of applying bichromatic ($q=2$) laser pulses for laser and plasma based electron acceleration. As discussed in our earlier work \cite{Pocsai-LPB}, within the confines of the present approach the $n_{m} < 1$ case agrees quite well with the $n_{m} = 1$ case, since at relevant plasma densities the refraction index of an underdense plasma differs negligibly from unity.

\begin{figure}[!h]
\begin{center}
	\includegraphics[scale=.4]{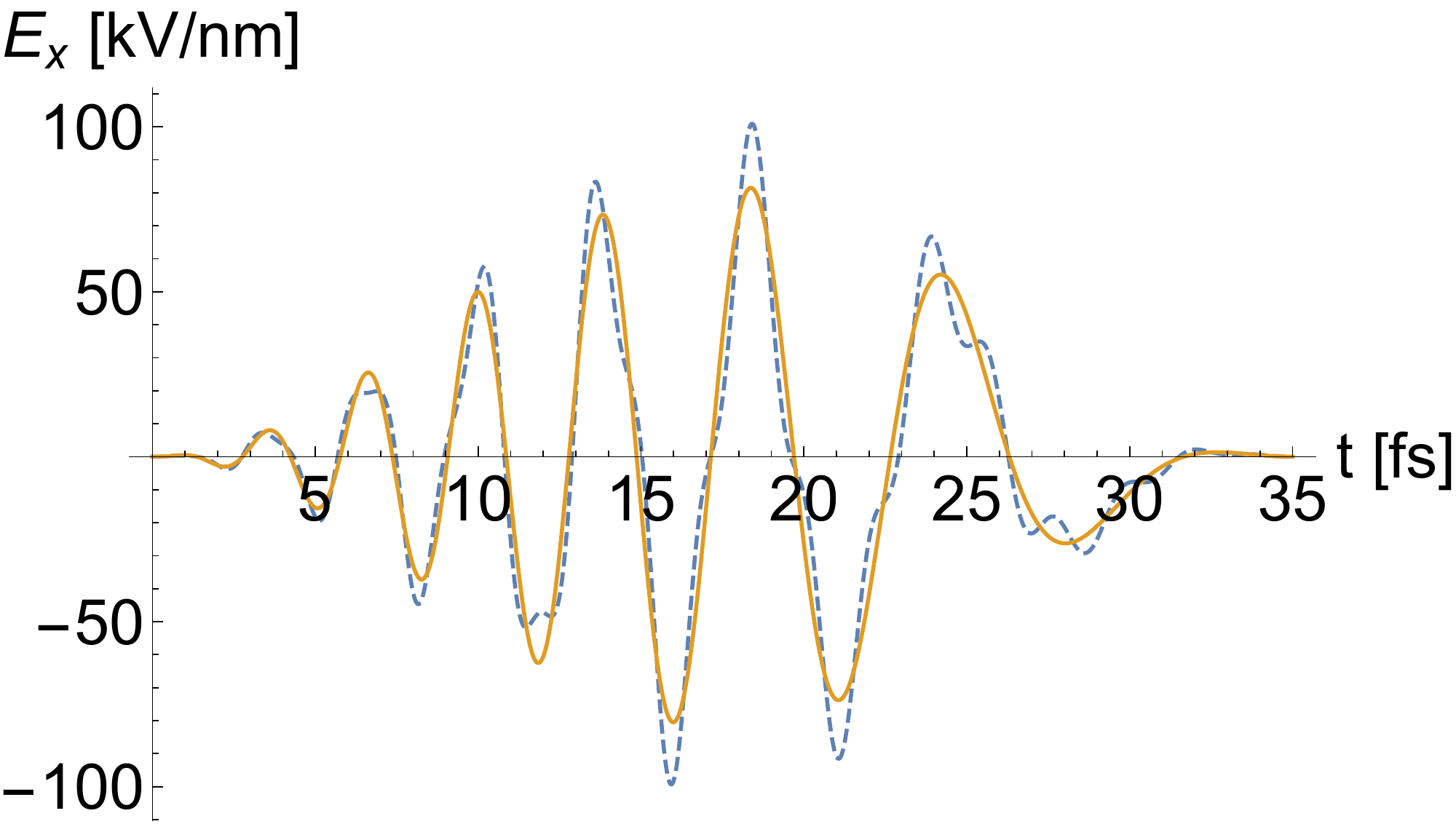}
	\caption{The $x$ component of the electric field of a chirped, monochromatic (solid line) and bichromatic (dashed line) plane wave pulse. $\lambda = 800 \, \mathrm{nm}$, $T = 35 \, \mathrm{fs}$, $a_{0} = 0.22$, $A = 0.48$, $\sigma_{1} = -5.510 \cdot 10^{-3}$, $\sigma_{2} = -1.311 \cdot 10^{-3}$, $\varphi_{1} = 0$. Note the enhancement of the accelerating field for the bichromatic case.}\label{fig:mhsh}
\end{center}
\end{figure}

We investigated the effect of the presence of the second harmonic. We found that by properly chosen parameters the energy gain of the electron can be enhanced by $4 \, \%$ or more. Graphically said, this is due to the increment of the accelerating field at the sharp rising and falling edges at the last oscillations of the electric field (see Fig.~\ref{fig:mhsh}).

In our calculations we chose the relative phase of the two harmonics to be zero. We compared the energy gain provided by a monochromatic and a bichromatic pulse such that the two pulses have the same intensity. This means that if the intensity parameter of a monochromatic pulse is $a_{0}$, then the intensity parameter of the corresponding bichromatic pulse is $b a_{0}$ with
\begin{equation}
	b = \parenth{\cfrac{\displaystyle\int_{\Theta_{i}}^{\Theta_{f}} \abs{\mathbf{E}_{1}}^{2} \diff \Theta}{\displaystyle\int_{\Theta_{i}}^{\Theta_{f}} \abs{\mathbf{E}_{1} + \cfrac{A}{2} \mathbf{E}_{2}}^{2} \diff \Theta}}^{1/2}
\end{equation}
with $\Theta_{i}$ and $\Theta_{f}$ being the plane wave arguments at times $t_{i}$ and $t_{f}$ as the interaction between the electron and the pulse starts and finishes. The subscripts $i$ and $f$ denote ``initial'' and ``final'', respectively. This normalization guarantees that the monochromatic and bichromatic pulses have the same intensity and validates the comparison of the energy gains from the two different pulses.

The additional energy gain via the presence of the second harmonic had been investigated in the following way. We took a monochromatic pulse with a fixed pulse duration at a given intensity and determined the optimal laser parameters and initial conditions that provide the most energy gain. After that we sought the optimal intensity ratio between the main and second harmonic while all the other parameters stayed fixed. At the determination of the optimal parameters we used the popular Nelder--Mead (also called ``downhill simplex'') method \cite{nelder_simplex_1965}.

At first we present the results for plane wave pulses, then for Gaussian pulses. We emphasize only the additional energy gain caused by the presence of the second harmonic. That is, we normalize the energy gains ($\Delta E$) to the optimal energy gain achieved by the corresponding a monochromatic pulse ($\Delta E_{0} \parenth{A}$) at the \emph{same intensity}. The relative energy gain is defined by $\Delta E / \Delta E_{0} (A)$. We also investigate the behavior of the energy gain as a function of the parameters of the two harmonics at fixed intensity ratio, namely, the chirp parameters and the carrier--envelope phase. In these cases we also scale the energy gain to unity, the normalization factor is the maximal energy gain (denoted by $\Delta E_{\textrm{max}}$) in a given parameter range. The scaled energy gain is also referenced as relative energy gain as is defined by $\Delta E / \Delta E_{\textrm{max}}$. To avoid ambiguity, the labels on the figures have been explicitly denoted. For a better transparency we note that the values of $a_{0} = 0.22, \, 7, \, 12$ and $22$ correspond to the intensities of $I = 10^{17}$, $10^{20}$, $3 \cdot 10^{20}$ and $10^{21} \, \mathrm{W} / \mathrm{cm}^{2}$, respectively.

\subsection{Plane wave pulses}
\begin{figure}
\begin{center}
	\includegraphics[scale=.55]{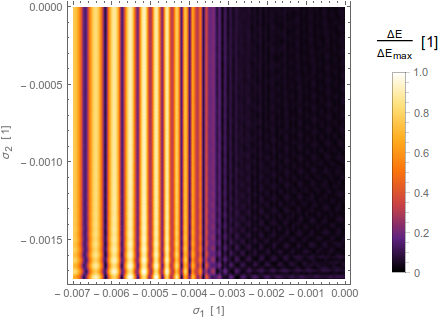}
	\caption{The relative energy gain as a function of the dimensionless chirp parameters. Note that the $\sigma_{2}$-dependence is negligible. $\lambda = 800 \, \mathrm{nm}$, $T = 75 \, \mathrm{fs}$, $a_{0} = 22$, $A = 0.7$, $\varphi_{1} = 0$.}\label{fig:pw-s1s2}
\end{center}
\end{figure}

We found that the net energy gain depends very weakly on the chirp parameter of the second harmonic. The energy gain is dominated by the chirp parameter of the main harmonic, this is indicated by the vertical stripes on Fig.~\ref{fig:pw-s1s2}. Practically this means that $\sigma_{2}$ can be chosen to be zero. As a consequence, both the experimental realization and numerical calculations are easier.

\begin{figure}
\begin{center}
	\includegraphics[scale=.42]{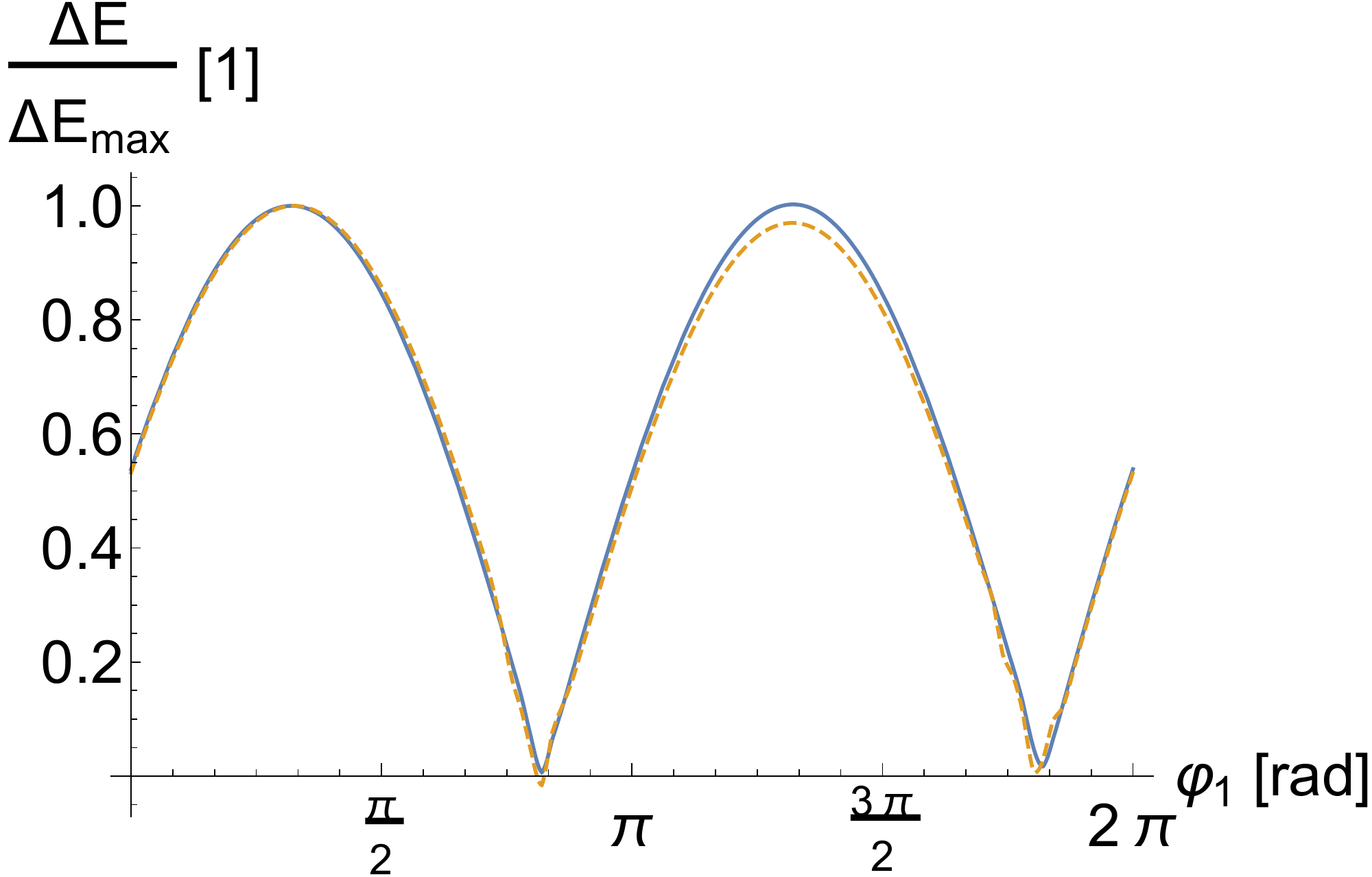}
	\caption{The relative energy gain as a function of the carrier--envelope phase. $\lambda = 800 \, \mathrm{nm}$, $T = 75 \, \mathrm{fs}$, $A = 0.7$, $\sigma_{1} = -5.518 \cdot 10^{-3}$, $\sigma_{2} = -1.732 \cdot 10^{-3}$, $a_{0} = 22$ (solid line), $a_{0} = 7$ (dashed line). Note the sensitivity to $\varphi_{1}$!}\label{fig:pw-phigain}
\end{center}
\end{figure}

\begin{figure}
\begin{center}
	\includegraphics[scale=.42]{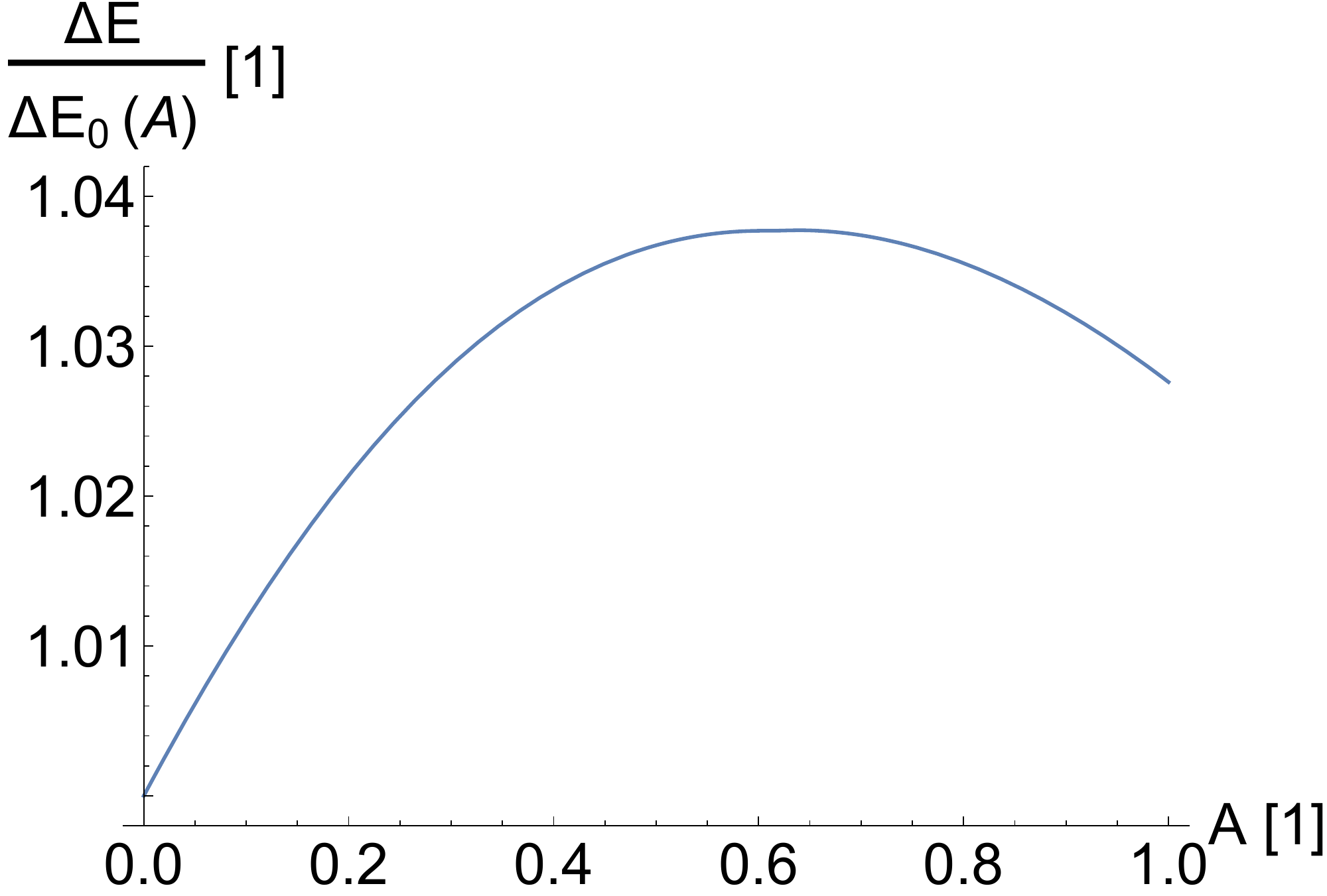}
	\caption{The relative energy gain as a function of the relative amplitude of the harmonics. The presence of the second harmonic enhanced the energy gain by about $4 \, \%$. $\lambda = 800 \, \mathrm{nm}$, $T = 75 \, \mathrm{fs}$, $a_{0} = 22$, $\sigma_{1} = -5.530 \cdot 10^{-3}$, $\sigma_{2} = -1.732 \cdot 10^{-3}$, $\varphi_{1} = 0$.}
	\label{fig:pw-again}
\end{center}
\end{figure}

It is not surprising that the carrier--envelope phase plays a very important role. However, it is not obvious that the energy gain depends very sensitively on $\varphi_{1}$: there are two ``worst-case'' values at which the net energy gain is zero. There are also two optima, near $\pi / 4$ and $3 \pi /4$ (see Fig.~\ref{fig:pw-phigain}).

Fig.~\ref{fig:pw-again} shows the key point of our study: at properly chosen laser parameters the energy gain of a single electron can be enhanced with approximately $4 \, \%$ by mixing the second harmonic to the main harmonic with a suitable intensity ratio. In the next subsection it can also be seen that if the intensity ratio has not been set correctly, then the application of the second harmonic results in a net energy loss compared to the monochromatic case (see Fig.~\ref{fig:gauss-again}).

\subsection{Gaussian pulses}

In general, a Gaussian pulse can transfer the most energy to a single electron if the electron initially moves on-axis an co-propagates with the beam. Namely, the initial momentum of the electron has the form of $\mathbf{p_{0}} = p_{0} \mathbf{e}_{z}$. This choice guarantees that the interaction length will be as high as possible. It is also important to initially place the electron far enough to not to feel the electric field of the pulse. The initial position of the electron has the form of $\mathbf{x}_{0} = x_{0} \mathbf{e}_{z}$. The dimensionless initial position is specified by $x_{0} = \pi \omega_{0} T$ that corresponds to a $\pi c T$ distance from the center of the pulse. After these considerations it should be clear that the larger the value of $p_{0}$ the larger the energy gain. This has been also confirmed by our numerical calculations, therefore we set $p_{0} = 2$ every time. This initial condition corresponds to $\gamma = \sqrt{5}$.

\begin{figure}
\begin{center}
	\includegraphics[scale=0.42]{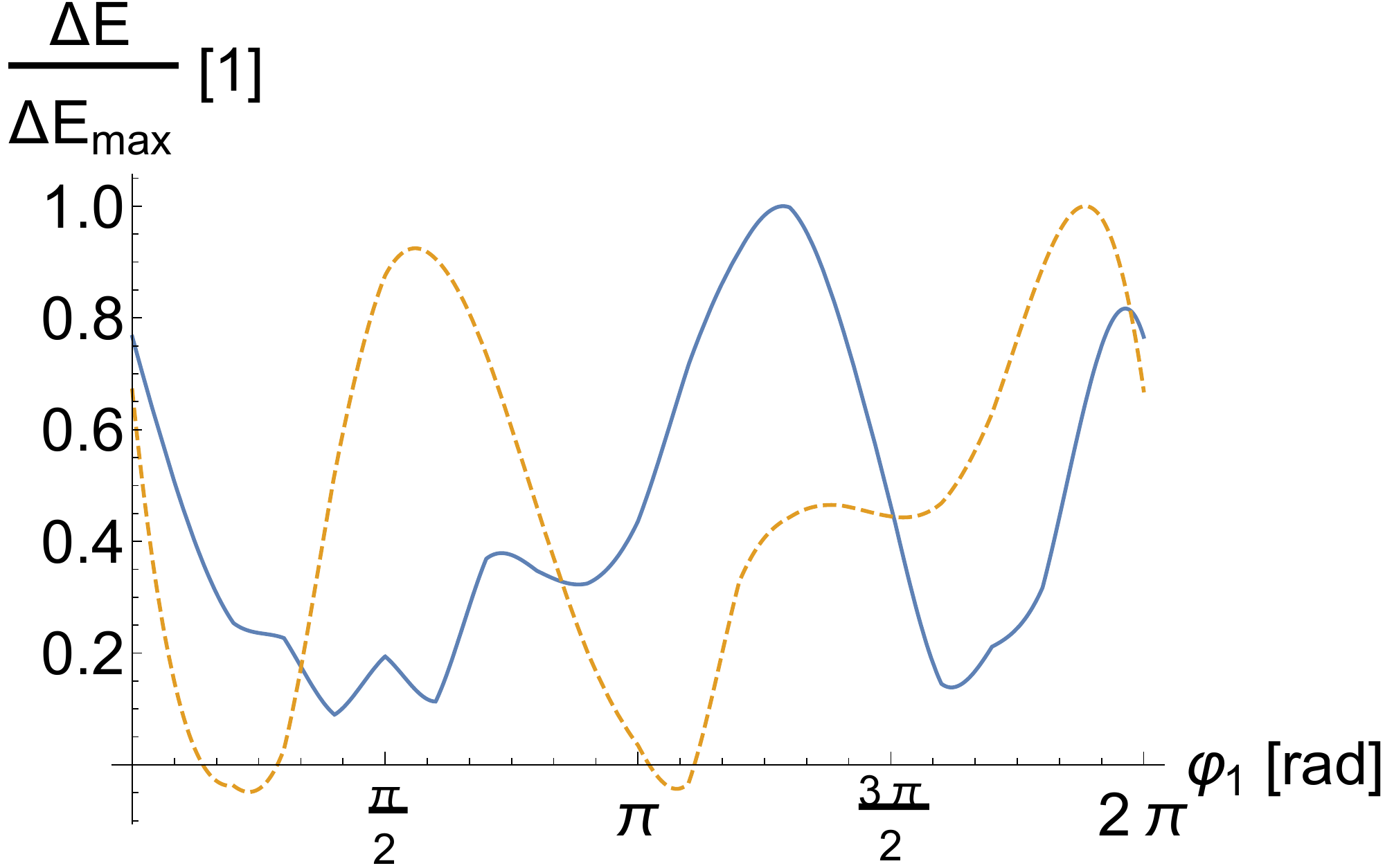}
	\caption{The relative energy gain as a function of the carrier--envelope phase. $\lambda = 800 \, \mathrm{nm}$, $T = 5 \, \mathrm{fs}$, $\sigma_{1} = 8.494 \cdot 10^{-3}$, $\sigma_{2} = 0$, $A=0.24$. $a_{0} = 7$, $W_{0} = 22.44 \pi$ (solid line), $a_{0} = 12$, $W_{0} = 20 \pi$ (dashed line).}\label{fig:gauss-cepgain}
\end{center}
\end{figure}

\begin{figure}
\begin{center}
	\includegraphics[scale=.35]{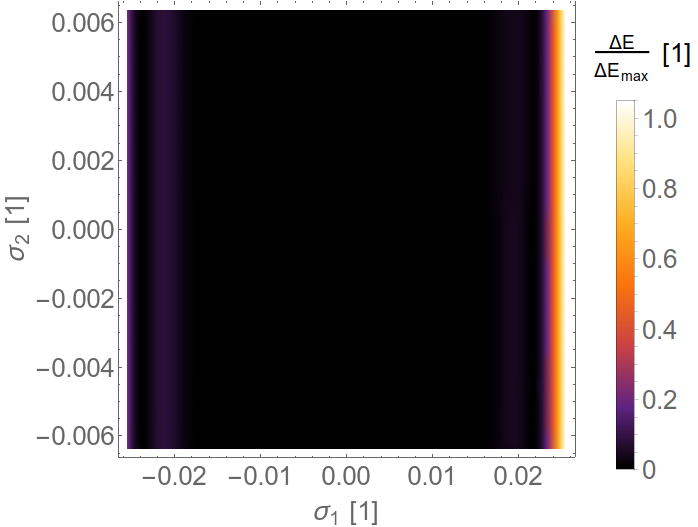}
	\caption{The relative energy gain as a function of the chirp parameters. $\Delta E$ depends very weakly on $\sigma_{2}$. $\lambda = 800 \, \mathrm{nm}$, $T = 5 \, \mathrm{fs}$, $a_{0} = 2$, $W_{0} = 20 \pi$, $A = 0.1$, $\varphi_{1} = 2.058$.}\label{fig:gauss-s1s2}
\end{center}
\end{figure}

For bichromatic Gaussian pulses we found that the application of the second harmonic is only reasonable at short pulse durations, namely in the $5 \, \mathrm{fs} - 15 \mathrm{fs}$ range. We analysed the energy of the electron as a function of time during the interaction with the beam and found that at usual pulse durations, e.g.~$35 \mathrm{fs} - 40 \, \mathrm{fs}$ or above, the second harmonic only causes a small, oscillatory perturbation in the energy--time function.

\begin{figure}
\begin{center}
	\includegraphics[scale=.42]{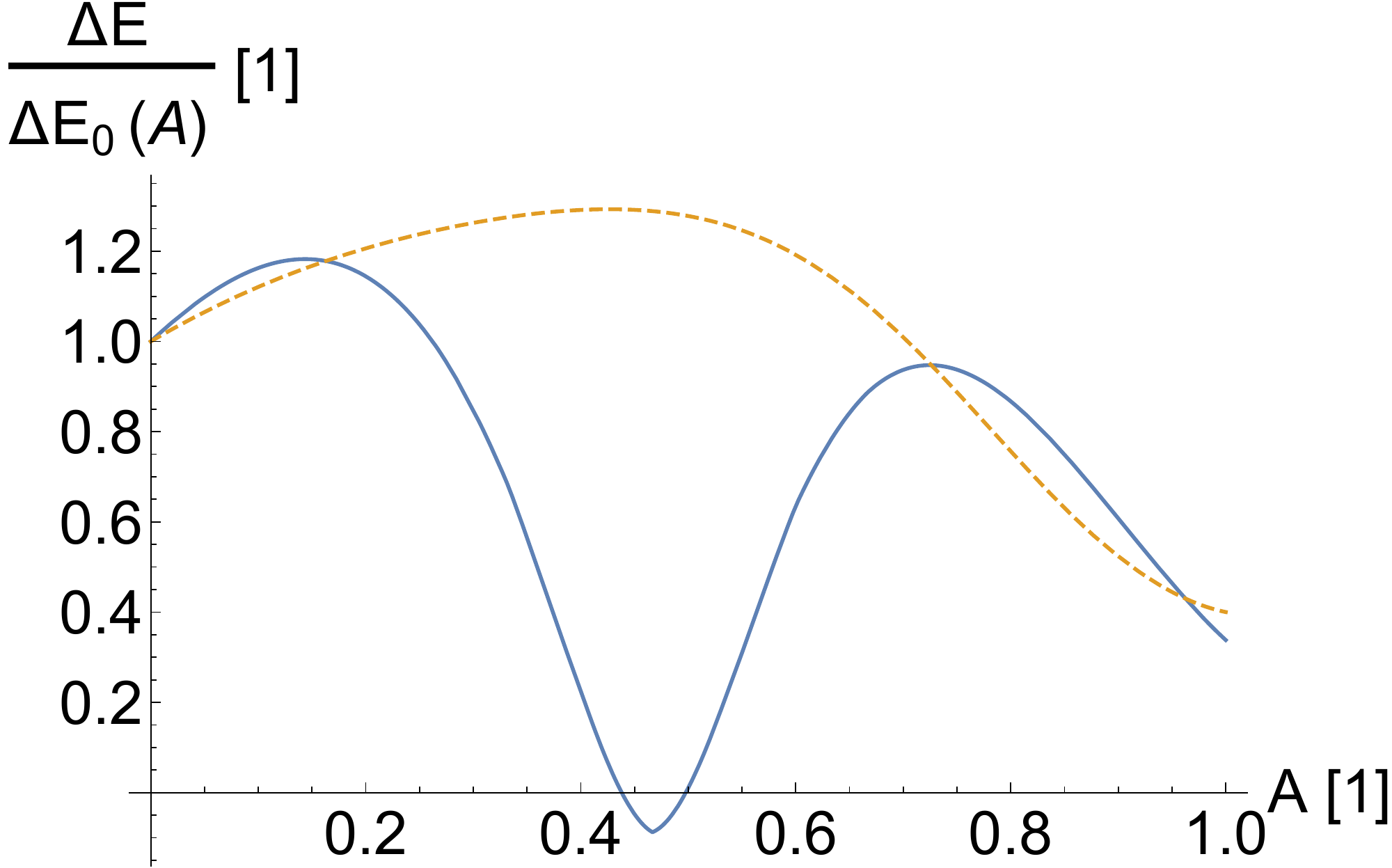}
	\caption{The relative energy gain as a function of the relative amplitude of the harmonics. The second harmonic enhanced the energy gain by about $20 - 30 \, \%$, compared to the monochromatic case. $\lambda = 800 \, \mathrm{nm}$, $T = 5 \, \mathrm{fs}$, $\sigma_{1} = 8.494 \cdot 10^{-3}$, $\sigma_{2} = 0$. $a_{0} = 7$, $W_{0} = 20 \pi$, $\varphi_{1} = 0$ (solid line), $a_{0} = 12$, $W_{0} = 22.44 \pi$, $\varphi_{1} = 0$ (dashed line).}
	\label{fig:gauss-again}
\end{center}
\end{figure}

However, at short pulse lengths, we found considerable additional gains due to the presence of the second harmonic. As expected, the energy gain depends very sensitively on the carrier--envelope phase. For some values of $\varphi_{1}$, the net energy gain can be even negative (see Fig.~\ref{fig:gauss-cepgain}).

The energy gain depends very weakly on the chirp parameter of the second harmonic, as presented on Fig.~\ref{fig:gauss-s1s2}, the dominant parameter---along with the initial momentum, carrier--envelope phase and beam waist---is the chirp parameter of the main harmonic.

As mentioned above, we found that the application of the second harmonic may result in a considerable additional energy gain. The enhancement may be $20 \, \%$, greater than conjectured, but it may reach even $30 \, \%$ (see Fig.~\ref{fig:gauss-again}). This promising result suggests that bichromatic laser pulses could be efficiently used for laser and laser--plasma based electron acceleration.

\section{Summary}
An effective theory for describing laser--plasma based electron acceleration has been presented. Earlier we showed that the background effects of the plasma can be incorporated into its refraction index. This way, the laser-driven plasma based acceleration can be well approximated with the pure laser-based acceleration, and the basic phenomena can be studied in a numerically and theoretically simple manner. In the present paper we investigated the acceleration mechanisms driven by bichromatic laser fields. We found that by properly chosen parameters the energy gain can be enhanced by $4 \, \%$ for plane wave pulses and even $30 \, \%$ for Gaussian pulses, compared to the monochromatic case. These are promising results that confirm that it is useful to apply bichromatic driver pulses for laser and laser--plasma based acceleration.
It would be interesting to perform the same calculations with such Gaussian pulse shapes that are exact solutions of Maxwell's equations and compare them with our most recent results. P.~Varga and P.~Török derived such solutions and found that at wide focusing---that is, if the beam diameter is greater than ten times the laser wavelength---the paraxial approximation provides satisfactory results, however, at tight focusing, the differences are significant \cite{varga_gaussian_1998}.

\section*{Acknowledgement}
S.V.~has been supported by the National Scientific Research Foundation OTKA, Grant No.~K 104260. Partial support by the ELI-ALPS Project is also acknowledged. The ELI-ALPS Project (GOP-1.1.1-12/B-2012-0001) is supported by the European Union and co-financed by the European Regional Development Fund. M.A.~Po\-cs\-a\-i has been supported by the ``Preparations for the concerned sectors for educational and R\&D activities related to the Hungarian ELI Project'', Grant No.~TAMOP-4.1.1.C-12/1/KONV-2012-0005.

%\section*{References}
\bibliographystyle{elsarticle-num-names}
\bibliography{references2}
\end{document}